\documentclass[prb, twocolumn, amssymb, amsmath,aps,superscriptaddress,showpacs]{revtex4}
\usepackage{natbib}
\usepackage{colordvi}
\usepackage[dvips]{color}
\usepackage{graphicx}
\usepackage{epstopdf}
\usepackage{bm}
\usepackage{amsmath, amsfonts}

\begin{document}

\title{Proximity effects in a topological-insulator/Mott-insulator heterostructure}
\author{Suguru Ueda}
\affiliation{Department of Physics, Kyoto University, Kyoto 606-8502, Japan}
\author{Norio Kawakami}
\affiliation{Department of Physics, Kyoto University, Kyoto 606-8502, Japan}
\author{Manfred Sigrist}
\affiliation{Theoretische Physik, ETH Z\"urich, CH-8093 Z\"urich}

\date{\today}

\begin{abstract}
We investigate proximity effects in a correlated heterostructure of a two-dimensional Mott insulator (MI) and a topological insulator (TI) by employing inhomogeneous dynamical mean-field theory. We show that the edge state of the TI induces strongly renormalized mid-gap states inside the MI region, which still have a remnant of the helical energy-spectrum. The penetration of low-energy electrons, which is controlled by the interface tunneling $V$, largely enhances the electron mass inside the MI and also splits a single Dirac-cone at edge sites into the spatially-separated two Dirac-cones in the strong $V$ region.
\end{abstract}

\maketitle


Remarkable progress has been made in the study of the topological insulator (TI) as a new class of materials characterized by an energy gap in the bulk but gapless edge (surface) states at its boundary~\cite{PhysRevLett.95.146802, bernevig2006quantum, PhysRevLett.98.106803}. In recent experiments, the observation of such TIs is reported for HgTe/CdTe quantum wells~\cite{bernevig2006quantum,konig2007quantum} and some bismuth compounds (such as Bi$_2$Se$_3$ and Bi$_{1-x}$Sb$_x$) ~\cite{zhang2009topological,xia2009observation,hsieh2008topological,hsieh2009observation}, corresponding to two- and three-dimensional TIs, respectively. Edge states of TIs are protected by non-trivial topological properties of the electronic bulk spectrum, thereby being robust against small perturbations conserving time-reversal symmetry, such as non-magnetic impurities~\cite{PhysRevLett.96.106401}. Hence, as long as the bulk gap exists, the low-energy physics of the TI is dominated by the edge states.

Heterostructures involving the TIs are currently the subject of intensive studies as they might have applications in future spintronics devices. They also provide a versatile platform for searching exotic interface phenomena, such as Majonara bound states~\cite{PhysRevLett.100.096407,PhysRevB.81.241310} and anomalous magnetoresistance~\cite{PhysRevB.81.121401}. In particular, we would like to explore the interface physics which emerges in a heterostructure composed of a TI and strongly correlated materials, since the previous studies of TI have mainly focused on heterostructures involving non-interacting electrons.
 
Regarding heterostructures with electron correlations, recent years have seen tremendous advances in producing high-quality interfaces composed of various materials such as band-insulator/Mott-insulator (BI/MI) or different types of band insulators. Unusual properties have been discovered at these interfaces, such as strongly confined metallic phases~\cite{ohtomo2002artificial}, magnetism~\cite{brinkman2007magnetic}, and superconductivity~\cite{reyren2007superconducting} to name a few. The occurrence of a metallic interface through electronic rearrangement is one of the intriguing features of the BI/MI heterostructures~\cite{okamoto2004electronic, PhysRevB.70.241104, PhysRevB.85.235112, PhysRevLett.108.117003, PhysRevLett.101.066802, PhysRevLett.108.246401}. Naturally we may ask how the situation is modified when the band insulator is replaced by a topological insulator (TI). The interface metallic state should be influenced by the presence of topological edge states and the interplay between such edge states and the strong electron correlation can give rise to novel physical properties. 

In this study, we analyze the electronic properties at a two-dimensional heterostructure consisting of a paramagnetic Mott insulator (MI) and a TI, by using a rather simple microscopic model for both insulators of TI/MI heterostructure considering onsite Coulomb repulsion but ignoring long-range interactions. The electronic correlations are treated by dynamical mean-field theory~\cite{RevModPhys.68.13} so that we can follow the renormalization of quasiparticles penetrating the MI and their interplay with the nearly localized degrees of freedom of the MI. We will elucidate how the important parameters such as the (onsite) Coulomb repulsion and the coupling (hopping) between the two insulators influence the metallic quasiparticle states.


The model describes a two-dimensional single-quantum-well geometry (square lattice) composed of a TI sandwiched by identical paramagnetic MIs on both sides. In view of the fact that our TI is based on a two-orbital model, introduced below, we choose also for the MI a configuration of two independent bands at half filling with strong Coulomb repulsion. 
The Hamiltonian is decomposed into $H = H_{\text{TI}} +\sum _{i=\text{R},\text{L}}(H_{\text{M}}^{i} +H_{\text{V}}^{i})$ with
%
\begin{eqnarray}
H_{\text{M}}^{\text{R},\text{L}} &=& \sum _{\langle i,j \rangle, \sigma, \alpha} t_{\alpha}
						\hat {c}^{\dagger}_{i\sigma \alpha}\hat{c}_{j\sigma \alpha} +U_{\text M}\sum _{i \alpha} \hat{n}_{i \uparrow \alpha}\hat{n}_{i \downarrow \alpha}, \label{eq:scm} \\
H_{\text{V}}^{\text{R},\text{L}} &=& \sum _{\langle i,j \rangle, \sigma, \alpha} V_{\alpha} \left (
						\hat{c}^{\dagger}_{i \sigma \alpha}\hat{a}_{j \sigma \alpha} +\hat{a}^{\dagger}_{i \sigma \alpha}\hat{c}_{j \sigma \alpha} \right ) \label{eq:hybridization}.
\end{eqnarray}
%
Here, $H_{\text{M}}^{\text{R}}$ ($H_{\text{M}}^{\text{L}}$) denotes the Hamiltonian for the MI on the right (left) edge of the TI region, and the coupling between these regions is implemented by the hybridization matrix, $H_{\text{V}}^{\text{R}}$ ($H_{\text{V}}^{\text{L}}$). The parameters $t_{\alpha}$ and $V_{\alpha}$  are the hopping integrals for orbital $\alpha$. We assume $t_1=-t_2=-t$ and $V_1=-V_2=-V$ for simplicity. The fermion operators $\hat c^{\dagger}_{i \sigma \alpha}$ and $\hat a^{\dagger}_{i \sigma \alpha}$ ($\hat c_{i \sigma \alpha}$ and $\hat a_{i \sigma \alpha}$) create (annihilate) a spin $\sigma=\uparrow, \downarrow $ electron of orbital $\alpha=1,2$ at site $i$ on the square lattice. Note that $\hat c_{i \sigma \alpha}$ and $\hat a_{i \sigma \alpha}$ operate on the orbitals for MI and TI, respectively.  

For the TI region we introduce a generalized Bernevig-Hughes-Zhang (BHZ) model~\cite{PhysRevB.85.165138}, given by $ H_{\text{TI}} = H_{\text{BHZ}} +U_{\text{TI}}\sum_{i \alpha} \hat{n}_{i \uparrow  \alpha}\hat{n}_{i \downarrow  \alpha}$, where
%
\begin{eqnarray}
H_{\text{BHZ}} = \sum _{i,\sigma,\alpha} \epsilon _{\alpha} \hat {n}_{i \sigma \alpha}
  +\sum _{\substack {\langle i,j \rangle,\\ \sigma, \alpha, \beta}}
   \hat {a}^{\dagger}_{i \sigma \alpha} \left [ \hat {t}_{\sigma}(\delta) \right ]_{\alpha \beta}
   \hat {a}_{j \sigma \beta}, \label{eq:bhz+u} \\
\hat {t}_{\sigma}(\pm x) = 
	\begin{pmatrix}
	t_1 \:\:\: \pm i \sigma t_{\text{so}} \\
	\pm i \sigma t_{\text{so}} \:\:\: t_2
	\end{pmatrix}, \:\:\:
\hat {t}_{\sigma}(\pm y) =
	\begin{pmatrix}
	t_1 \:\:\: \pm t_{\text{so}} \\
	\mp t_{\text{so}} \:\:\: t_2
	\end{pmatrix}.
\end{eqnarray}
%
Here, $\hat t_{\sigma}(\delta)$ with $\delta = \pm x$ ($\pm y$) denotes the spin dependent hopping integral along the $x$ ($y$)-direction. We assume $\epsilon _1 = -\epsilon _2 =-t$ in the following. The topologically non-trivial phase is driven via the finite inter-site and inter-orbital hybridization $t_{\text{so}}$ for $0 < |\epsilon _1| < 4t$. 

For our calculation we set the width of the TI (MI) region  to $20$ ($10$) unit cells along the $y$-direction, while the system keeps the translation symmetry in the $x$-direction. Unless otherwise mentioned, $t_{\text{so}}$ and $V$ are fixed to $0.25t$ and $t$, respectively, and the spin index $\sigma$ is dropped. Throughout this paper, we restrict ourselves to zero temperature and half filling ($\langle \hat {n}_{i} \rangle = \sum_{\sigma \alpha} \langle \hat {n}_{i \sigma \alpha} \rangle = 2$). 

We note here that correlation effects in TIs have been studied extensively, also suggesting that the strong electron correlation disturbs the topological nature~\cite{PhysRevLett.106.100403,PhysRevLett.107.010401,PhysRevB.85.165138} and even induces topological phases without gapless edge states~\cite{PhysRevB.83.205122}. Thus we may ask how the correlation effects manifest themselves at TI/MI interfaces. This is the central issue in the present paper.

In the following, many-body effects in the above Hamiltonian are treated within the inhomogeneous dynamical mean-field theory (IDMFT)~\cite{PhysRevLett.101.066802, PhysRevB.59.2549, PhysRevB.70.241104,PhysRevB.79.045130}, which solves the $y$-dependent self-energy as a diagonal matrix with self-consistent equations,
%
\begin{equation}
\mathcal {G}_{0\alpha \beta}^{-1}(y,y';\omega ) = 
	\left [ \int \frac{d k_x}{2\pi} G(y,y';k_x, \omega ) \right ]_{\alpha \beta}^{-1}
	+\Sigma_{\alpha \beta}(y; \omega ), \notag
\end{equation}
%
where, $G_{\alpha \beta}(y,y';k_x, \omega )$ and $\mathcal {G}_{0\alpha \beta}(y,y';\omega )$ are the lattice and cavity Green's functions, respectively. Note that we treat this system as a spatially modulated two-dimensional system with a finite strip width, following the treatment done in Ref.~\onlinecite{PhysRevB.85.165138}.

The local self-energy for the quantum impurity model is obtained using the exact diagonalization methods~\cite{PhysRevLett.72.1545}, suitably extended for the IDMFT analysis~\cite{PhysRevB.79.045130}. To avoid time-consuming numerics, we neglect the inter-band self-energy, $\Sigma_{12}(y; \omega)$ and $\Sigma_{21}(y; \omega)$. The validity of this approximation has been discussed in Ref.~\onlinecite{PhysRevB.85.165138}. Here, we employ the Lanzcos algorithm to solve the local Green's function, and the bath parameters for the finite-size system are obtained by minimizing the following function~\cite{liebsch2011temperature}:
%
\begin{equation}
\chi (y) = \sum _{\omega _n} | \, {\mathcal {G}_{0}^{\text{fs}}}(y; i\omega _n) -{ \mathcal{G} }_{0}(y, y; i\omega _n)|^2,
\end{equation}
%
where, $\mathcal { {G} }_{0}^{\text{fs}}(y; \, i\omega _n)$ is the non-interacting  Green's function of the impurity model for the discretized Matsubara frequency $\omega_n = (2n+1)\pi/\tilde {\beta}$ on chain $y$. We choose the number of the bath levels coupled to each band to $n_{\text b}=7$, and fix the inverse imaginary temperature $\tilde {\beta}=200$.


%
\begin{figure}[!t]
\centering
\includegraphics[width=0.85\linewidth]{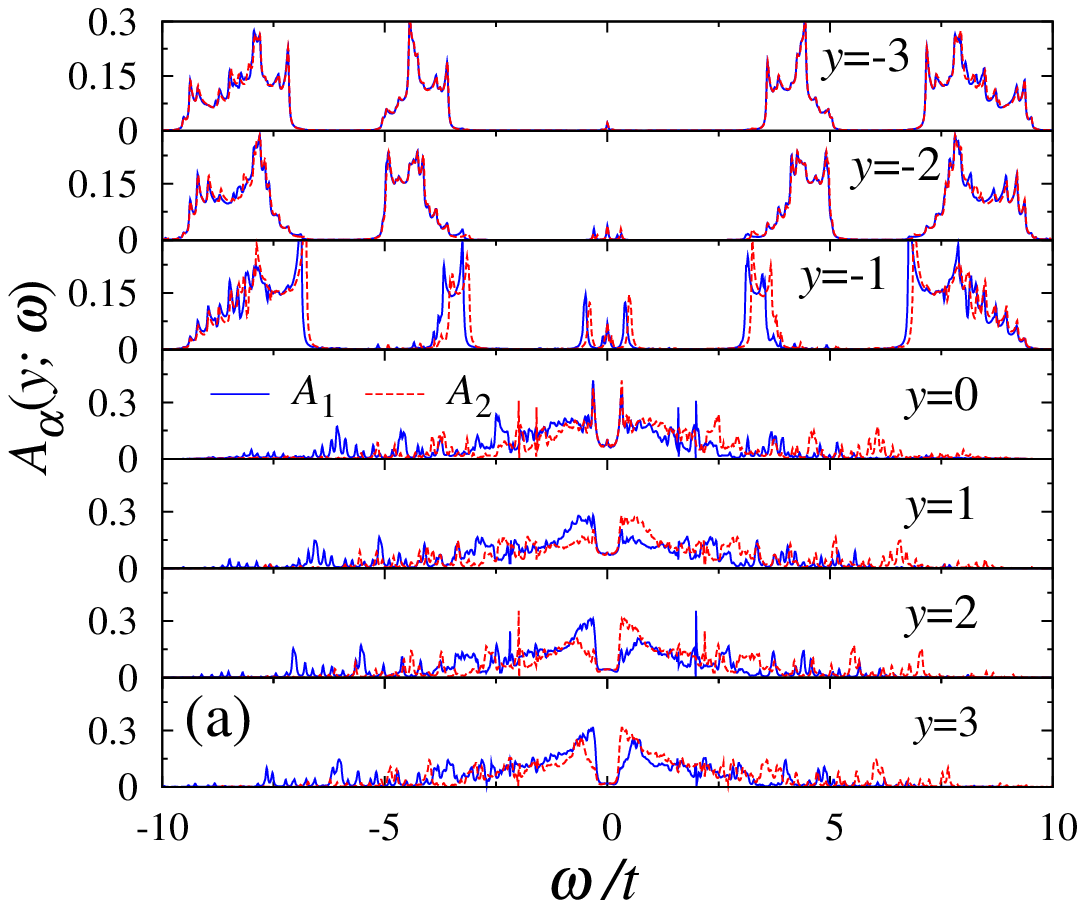}
\hfil
\includegraphics[width=0.85\linewidth]{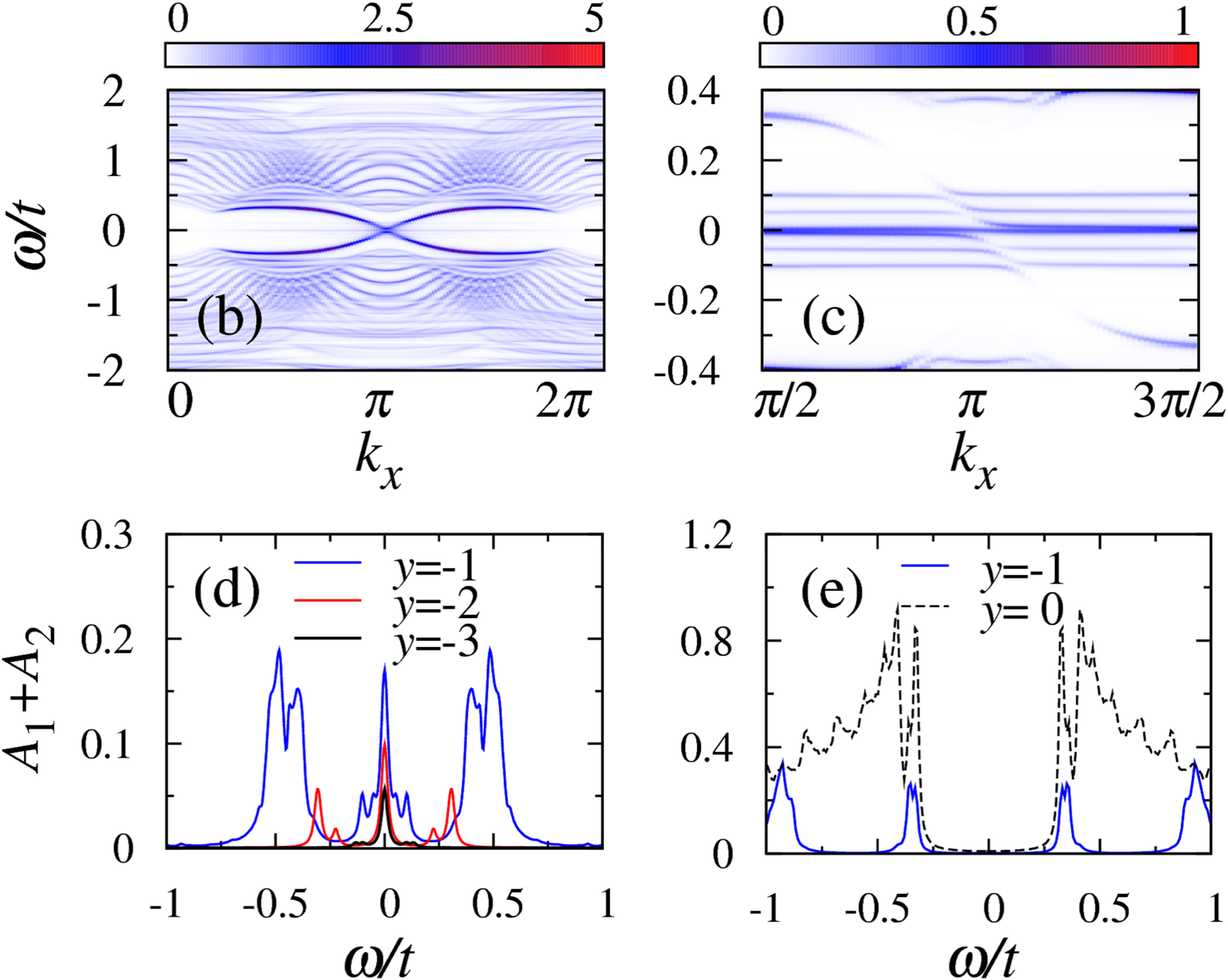}
\vspace{-1mm}
\caption{(a) The orbital-resolved local spectral-function $A_{\alpha}(y; \omega)$ for MI region (left half of the system) with $U_{\text M}=13.3t$ and $U_{\text {TI}}=4t$. The TI (MI) regions correspond to $y \geq 0$ ($y\leq -1$), and the solid and dashed lines indicate $A_{\text 1}$ and $A_{\text 2}$, respectively. Plots of the corresponding momentum-resolved spectral functions: (b) at $y=0$ (TI-edge) with both spin states and (c) at $y=-1$ (MI-edge) restricted to up-spin state only. (d) The spectral function ($A_{\text 1}+A_{\text 2}$) around zero frequency inside the MI region near $y=0$. The curves from top to bottom in the vicinity of zero frequency correspond to the spectral functions at $y=-1$, $-2$ and $-3$, respectively. (e) The same as in (d) for the interface of a trivial BI and MIs for $y=-1$ (solid line) and $y=0$ (dotted line).}
\label{fig:spec-func}
\end{figure}
%
%
\begin{figure}[b]
\centering
\includegraphics[width=0.85\linewidth]{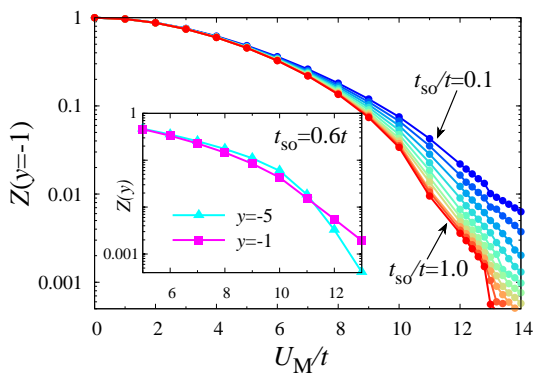}
\vspace{-3mm}
\caption{The quasi-particle weight $Z(y)$ at $y=-1$ as a function of $U_{\text M}$ with $t_{\text{so}}/t = 0.1, 0.2, \dots 1.0$ (increasing from top to bottom), fixing $Z(9)= 0.8\pm 0.001$. Inset: comparison of the corresponding $Z(y)$ at $y=-5$ (triangle) and $y=-1$ (square) for $t_{\text {so}}=0.6t$. }
\label{fig:mod-z}
\end{figure}
%

Figure~\ref{fig:spec-func}(a) shows the orbital-resolved local spectral function $A_{\alpha}(y; \omega) = -(1/\pi) \text{Im} G_{\alpha \alpha}(y, y; \omega +i\delta)$ for the MI region with $U_{\text M}=13.3t$ and $U_{\text {TI}}=4t$. Note that the strength of $U_{\text M}$ is slightly larger than the critical value $U_{\text c} \sim 13.2t$ for Mott transition in the bulk, while the strength of $U_{\text {TI}}$ is small enough to realize the non-trivial topological phase inside the heterostructure ($0\leq y \leq 19$)~\cite{PhysRevB.85.165138}. In this figure, we follow the evolution of the edge state in the energy gap when approaching the TI-edge ($y=0$) from $y=3$, whose existence is characteristic of the interface between topologically trivial and nontrivial materials. Actually the momentum-resolved spectral function, $A(y; k_x, \omega) = -(1/\pi) \sum_{\alpha} \text{Im}G_{\alpha \alpha}(y,y;k_x, \omega +i\delta)$, displays the edge state at $y=0$  in Fig.~\ref{fig:spec-func} (b). The edge state penetrates also into the MI region and induces a narrow band of mid-gap states. Figures~\ref{fig:spec-func} (c) and (d) show the momentum-resolved spectral function for up-spin states and the quasi-particle peak around $\omega\sim 0$ for $y\leq -1$, respectively. While the width of quasi-particle peak rapidly decreases away from $y=0$ in Fig.~\ref{fig:spec-func} (b), the existence of the renormalized quasiparticle states in the depth of the MI region is confirmed through the non-vanishing renormalization factor $Z(y)\equiv \left [ 1- \partial_{\omega} \text{Im} \Sigma(y; i\omega) \right ]_{\omega=0}^{-1}$. In addition, the energy spectrum describes the antisymmetric behavior across $k_x=\pi$, which implies that a heavy-fermion-like mid-gap state for $y\leq -1$ is induced by the penetration of the helical edge state. This scenario is further underlined by Fig.~\ref{fig:spec-func} (e), where the TI is replaced by the trivial BI with the same gap-size as the present TI. In this case, the heavy quasiparticles in the MI no longer exist, in contrast to the MI/TI interface. We also confirm that once the topological order is destroyed via the strong on-site Coulomb repulsion $U_{\text{TI}}$~\cite{PhysRevB.85.165138}, the spectral-weight for the mid-gap state simultaneously goes to zero (not shown here).

The formation of renormalized mid-gap states is understood in terms of a ``Kondo-type screening" mechanism between spin states in the MI and helical edge states. In the MI region, spin excitations are gapless, which appear as essentially localized free spins in the DMFT treatment. Such free spins are screened by helical edge states forming a Kondo-type resonance and establishing strongly renormalized mid-gap states in the first few layers of the MI. In this sense, the MI/TI interface effectively realize a Kondo lattice system with helical conduction electrons. Similar proximity effects can also be found in the studies of the MI/metal heterostructure~\cite{PhysRevLett.101.066802,PhysRevB.81.115134}. However, the distinctive difference becomes visible in $Z(y)$ around the interface. In the present case, the Kondo-type screening is driven only by helical edges states, so that the resulting mid-gap electron states become much heavier than that for MI/metal interfaces. It originates from the band-insulating nature of the TI with the energy gap $\Delta_{ \text{SO}}$, because in inhomogeneous correlated systems, the electron renormalization occurs strongly on their surface due to the reduced coordination number~\cite{PhysRevB.59.2549,PhysRevB.81.115134}. Hence, it is naturally expected that the renormalization of the mid-gap state is closely related to the gap size $\Delta_{ \text{SO}} \sim 4Zt_{ \text{so} }$~\cite{PhysRevB.85.165138}. Figure~\ref{fig:mod-z} in fact shows the monotonic suppression of $Z(y=-1)$ against the increase in $t_{ \text{so}}$. The inset of this figure also confirms the strong correlation effects around the interface for weak $U_{\text M}$, while this behavior is inverted for large $U_{\text M}$ where the penetration of the edge state mainly controls the spatial modulation of $Z(y)$. 
%
\begin{figure}[!t]
\centering
\includegraphics[width=0.85\linewidth]{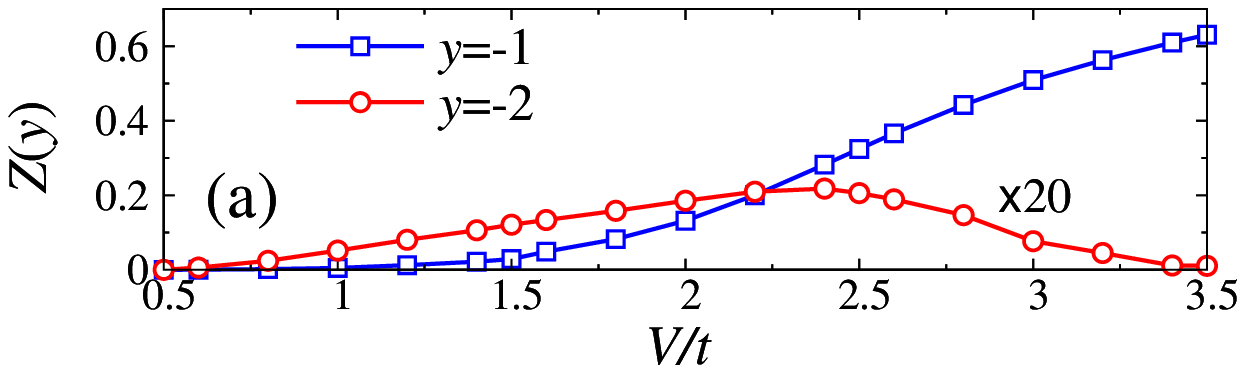}
\hfil
\includegraphics[width=0.88\linewidth]{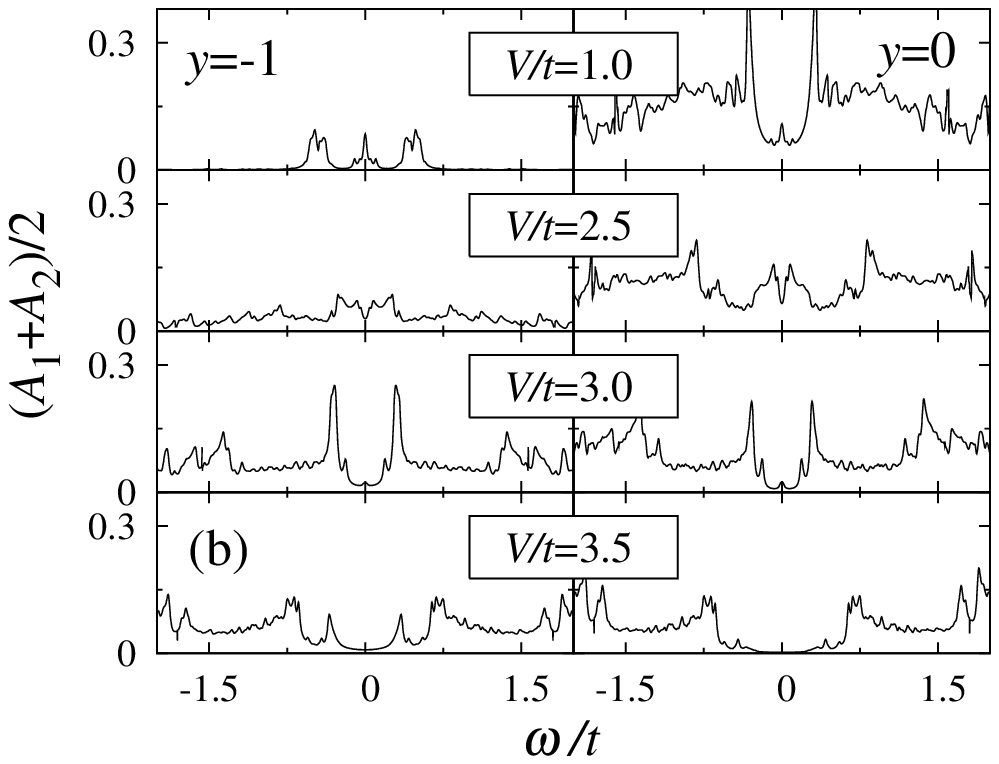}
\vspace{-3mm}
\caption{ (a) The $V$ dependence of the renormalization factor $Z(y)$ with $U_{\text M}=13.3t$ and $U_{\text {TI}}=4t$. The square and circle symbols represent the $Z(y)$ at $y=-1$ and $y=-2$, respectively; we note that $Z(-2)$ is enlarged 20 times. (b) The corresponding local spectral function: the left (right) panel for $y=-1$ ($y=0$).}  
\label{fig:eff-of-v-disp}
\end{figure}
%
%
\begin{figure}[!t]
\centering
\includegraphics[width=0.90\linewidth]{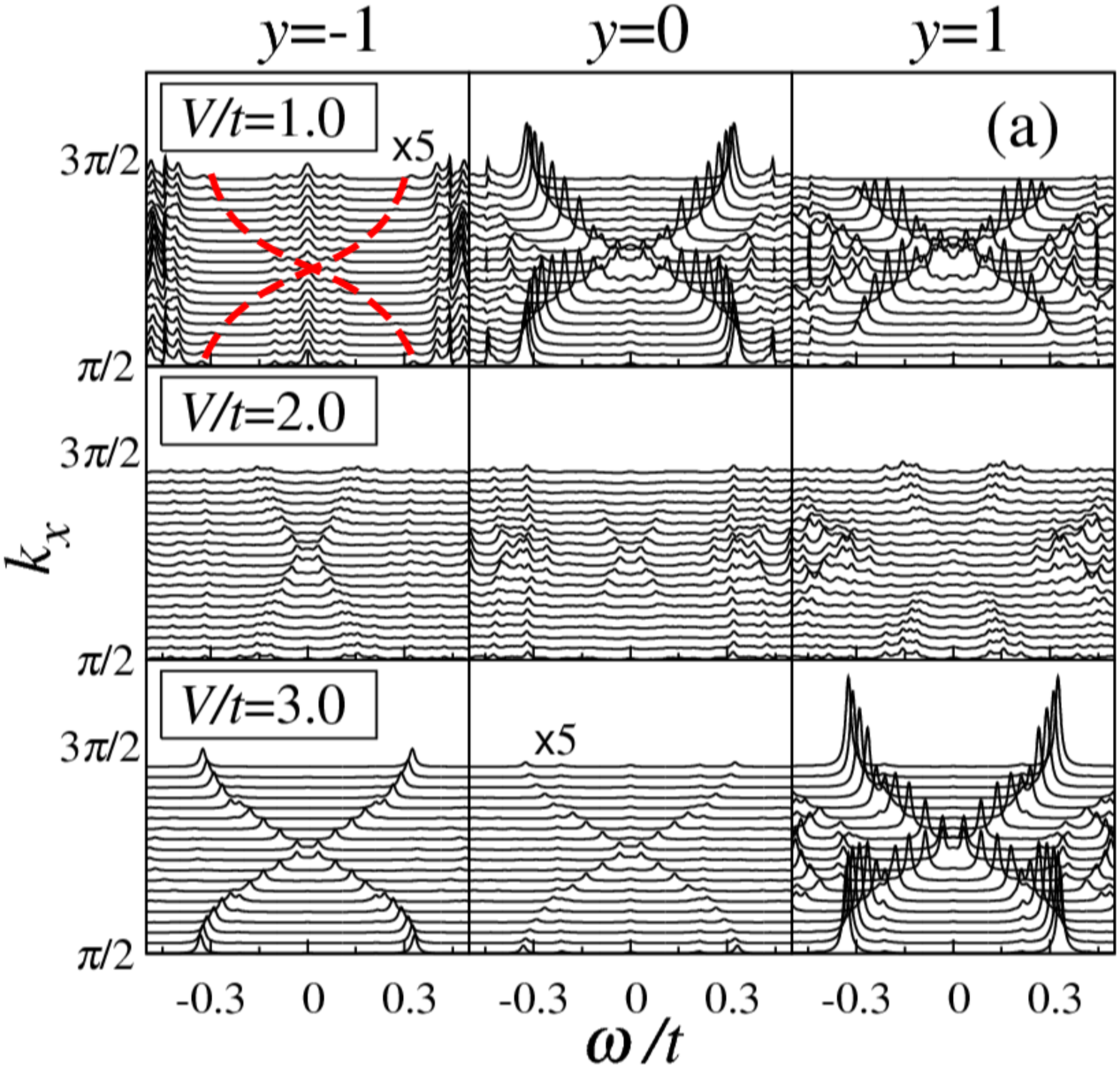}
\vspace{-2mm}
\includegraphics[width=0.90\linewidth]{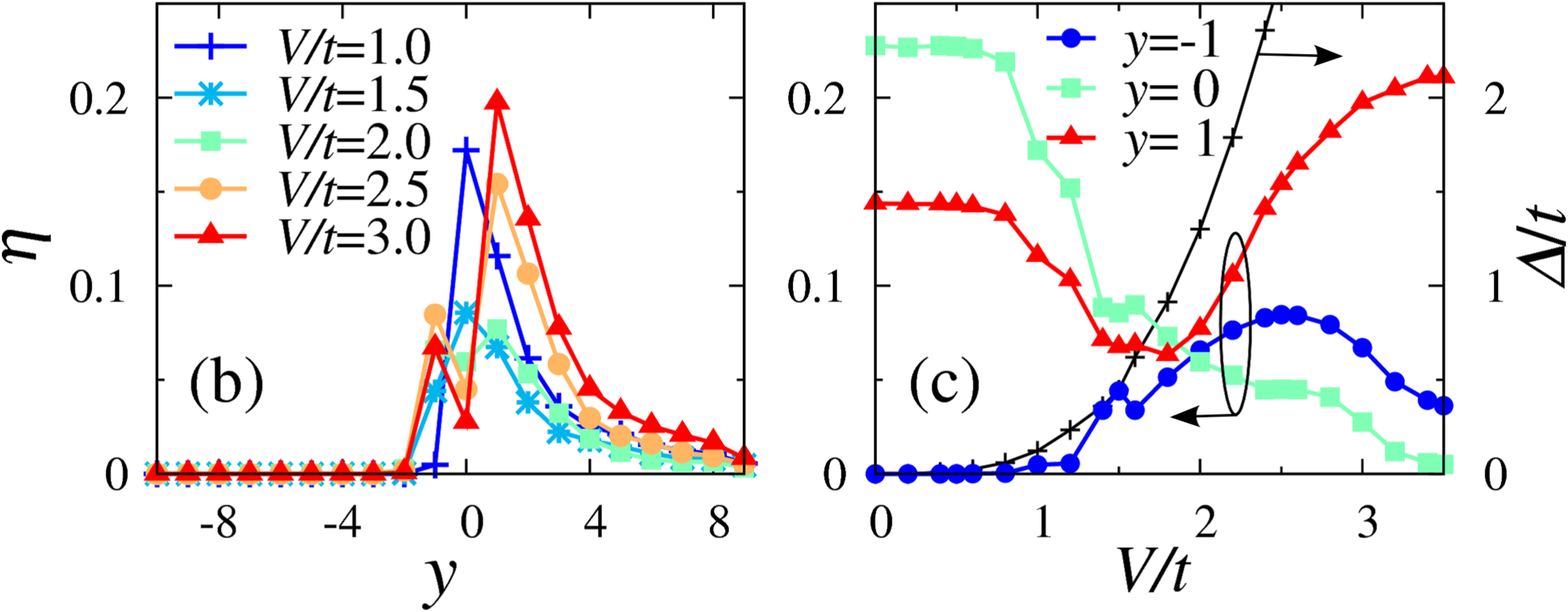}
\caption{(a) The momentum-resolved spectral function $A(y; k_x,\omega)$ for $y=-1$ (left), $y=0$ (center) and $y=1$ (right) with $U_{\text M}=13.3t$, $U_{\text{TI}}=4t$. Here $A(-1; k_x,\omega)$ for $V/t=1.0$ (left top) and $A(0; k_x,\omega)$ for $V/t=3.0$ (center bottom) are enlarged 5 times. (b) The corresponding helicity $\eta (y)$ defined in Eq.~\ref{helicity} as a function of $y$ with several values of $V/t$. (c) The $V$dependence of $\eta (y)$ for $y=0, \pm1$ (left axis) and the dimerization gap $\Delta \equiv 2V\sqrt {Z(-1)Z(0)}$ (right axis).} 
\label{fig:mom-spec}
\end{figure}
%

To get further insight into the mid-gap state, we now focus on the effect of interface electron tunneling $V$ between the TI and MI regions. We present the $V$-dependence of the renormalization factor in Fig.~\ref{fig:eff-of-v-disp} (a), where the change in the slope of $Z(-2)$ is found, in spite of the monotonic evolution of $Z(-1)$. This correlation enhancement for $y=-2$ can be understood as the band reconstruction at the interface: the formation of a dimerized state between $y=-1$ and $y=0$, whose energy gap is roughly given by $\Delta \sim 2V\sqrt{Z(-1)Z(0)}$~\cite{PhysRevLett.95.066402,PhysRevB.73.245118}. In the present system, $\Delta$ exceeds the band width of the edge state ($\sim \Delta_{\text{SO}} \simeq t$) above $V/t\sim 1.8$, which explains the $V$-dependence of $Z(-2)$ in Fig.~\ref{fig:eff-of-v-disp} (a) [see also Fig.~\ref{fig:mom-spec} (c)]. In Fig.~\ref{fig:eff-of-v-disp} (b), we further show the evolution of the energy gap around $\omega \sim 0$ for $y=-1$ and $y=0$. As anticipated, in both cases the gap structures are formed around $V/t \sim 2.5$ and the gap size monotonically grows with increasing $V$. 

An important question is how the topological edge state at $y=0$ depends on the dimerization between $y=-1$ and $y=0$. To this end, in Fig.~\ref{fig:mom-spec} (a), $A(y; k_x, \omega)$ for $y=0, \pm 1$ is plotted along the $k_x = \pi/2$ to $3\pi/2$. From the result we see that the spectral weight for the edge sites of the TI ($y=0$) is gradually suppressed as $V$ increases. On the contrary, the magnitude of $A(1; k_x, \omega)$ shows an upward turn across $V/t \sim 2.0$, and above $V/t \sim 3.0$ we find the characteristic $k_x$ dependence of $A(1; k_x, \omega)$ which is identical to that of $A(0; k_x, \omega)$ at $V/t=0$. Thus we conclude that in the limit $V \rightarrow \infty$, the edge state shifts its position from $y=0$ to $y=1$.

More closely looking at the bottom panels of Fig.~\ref{fig:mom-spec} (a), we find that the edge state exhibits an anomalous $y$-dependence: due to electron tunneling along $y$-direction, the magnitude of the Dirac-cone dispersion is expected to monotonically decrease away from $y=1$, but the obtained spectral-weight at $y=-1$ even exceeds that at $y=0$. This indicates that there exists two Dirac cones at $y = \pm 1$. Since the edge state should be localized at $y=1$ in the strong $V$ limit, we refer to the Dirac cone at $y=-1$, which looks like a copy of the edge state at $y=1$, as a {\it topological shadow edge-state} (TSE). 

We can characterize the edge states in the TSE by introducing the helicity function $\eta (y)$ defined as,
%
\begin{equation}
\eta (y) = \int _{ |\omega |<\Delta_{\text{SO}} } \frac{ d\omega d k_x}{ (2\pi )^2 } | A_{\uparrow}(y; \omega, k_x) -A_{\downarrow}(y; \omega, k_x)|, \label{helicity}
\end{equation}
%
where $A_{\sigma}$ is the spectral function ($A_1+A_2$) with spin $\sigma$ and the $\omega$-integral is limited within the gap $\Delta _{\text {SO}} = 4Zt_{\text{so}}$. We emphasize that $\eta (y)$ in Figs.~\ref {fig:mom-spec} (b) and~\ref {fig:mom-spec} (c) shows the good correspondence to the edge-state behavior in Fig.~\ref {fig:mom-spec} (a). Until the dimerization gap opens for $V/t \sim 2.5$, the magnitude of $\eta(-1)$ increases while that of $\eta(0)$ decreases with increasing V, as shown in Fig.~\ref {fig:mom-spec} (c) [see also Fig.~\ref{fig:eff-of-v-disp}(a)]. This may be understood in terms of the topological band-reconstruction. For relatively larger values of $V/t \lesssim 2.5$, the energy spectrums at $y=-1$ and $y=0$ are strongly hybridized with each other, which extends the topologically non-trivial band-structure toward $y=-1$. In this sense, the sites at $y=-1$ rather than $y=0$ shall be regarded as the edge of the TI. We note that when $V$ further increases, the edge state should shift its position from $y=-1$ to $y=1$. Therefore, the TSE for $V/t \sim 3$ can be understood as the remnant of this displacement, as plotted in Fig.~\ref{fig:mom-spec} (c).

To summarize, we presented the DMFT study of a minimal model for the heterostructure of the two-dimensional TI embedded in the MIs. Our results showed that the helical edge state at the end of the TI penetrates into the MI, even if the Hubbard gap is very large. We clarified that such proximity effects induce the strongly renormalized mid-gap state having a remnant of the helical energy spectrum inside the MI region. It was found that the correlation effects around the interface are strongly enhanced due to the spin-orbit gap in the TI region. We also demonstrated how the hybridization between the TI and MI affects the electron penetration, and found the enhanced correlation effect and the existence of the TSE inside the MI, driven by the band reconstruction at the interface. 

We acknowledge financial support by a Grant-in-Aid for the Global COE Program ``The Next Generation of Physics, Spun from Universality and Emergence" from MEXT of Japan. N.K. is supported by KAKENHI (No.20102008) and JSPS through its FIRST Program. S.U. is supported by a JSPS Fellowship for Young Scientists and M.S. is grateful for support by the Swiss Nationalfonds and the NCCR MaNEP.


\begin{thebibliography}{10}

\bibitem{PhysRevLett.95.146802}
C.~L. Kane and E.~J. Mele, Phys. Rev. Lett. {\bf 95},  146802  (2005).

\bibitem{bernevig2006quantum}
B. Bernevig, T. Hughes, and S. Zhang, Science {\bf 314},  1757  (2006).

\bibitem{PhysRevLett.98.106803}
L. Fu, C.~L. Kane, and E.~J. Mele, Phys. Rev. Lett. {\bf 98},  106803  (2007).

\bibitem{konig2007quantum}
M. K{\"o}nig {\it et~al.}, Science {\bf 318},  766  (2007).

\bibitem{zhang2009topological}
H. Zhang {\it et~al.}, Nature Physics {\bf 5},  438  (2009).

\bibitem{xia2009observation}
Y. Xia {\it et~al.}, Nature Physics {\bf 5},  398  (2009).

\bibitem{hsieh2008topological}
D. Hsieh {\it et~al.}, Nature {\bf 452},  970  (2008).

\bibitem{hsieh2009observation}
D. Hsieh {\it et~al.}, Science {\bf 323},  919  (2009).

\bibitem{PhysRevLett.96.106401}
C. Wu, B.~A. Bernevig, and S.-C. Zhang, Phys. Rev. Lett. {\bf 96},  106401
  (2006).

\bibitem{PhysRevLett.100.096407}
L. Fu and C.~L. Kane, Phys. Rev. Lett. {\bf 100},  096407  (2008).

\bibitem{PhysRevB.81.241310}
T.~D. Stanescu, J.~D. Sau, R.~M. Lutchyn, and S. Das~Sarma, Phys. Rev. B {\bf
  81},  241310  (2010).

\bibitem{PhysRevB.81.121401}
T. Yokoyama, Y. Tanaka, and N. Nagaosa, Phys. Rev. B {\bf 81},  121401  (2010).

\bibitem{ohtomo2002artificial}
A. Ohtomo, D. Muller, J. Grazul, and H. Hwang, Nature {\bf 419},  378  (2002).

\bibitem{brinkman2007magnetic}
A. Brinkman {\it et~al.}, Nature materials {\bf 6},  493  (2007).

\bibitem{reyren2007superconducting}
N. Reyren {\it et~al.}, Science {\bf 317},  1196  (2007).

\bibitem{okamoto2004electronic}
S. Okamoto and A. Millis, Nature {\bf 428},  630  (2004).

\bibitem{PhysRevB.70.241104}
S. Okamoto and A.~J. Millis, Phys. Rev. B {\bf 70},  241104  (2004).

\bibitem{PhysRevB.85.235112}
S. Ueda, N. Kawakami, and M. Sigrist, Phys. Rev. B {\bf 85},  235112  (2012).

\bibitem{PhysRevLett.108.117003}
K. Michaeli, A.~C. Potter, and P.~A. Lee, Phys. Rev. Lett. {\bf 108},  117003
  (2012).

\bibitem{PhysRevLett.101.066802}
R.~W. Helmes, T.~A. Costi, and A. Rosch, Phys. Rev. Lett. {\bf 101},  066802
  (2008).

\bibitem{PhysRevLett.108.246401}
A. Euverte {\it et~al.}, Phys. Rev. Lett. {\bf 108},  246401  (2012).

\bibitem{RevModPhys.68.13}
A. Georges, G. Kotliar, W. Krauth, and M.~J. Rozenberg, Rev. Mod. Phys. {\bf
  68},  13  (1996).

\bibitem{PhysRevB.85.165138}
Y. Tada {\it et~al.}, Phys. Rev. B {\bf 85},  165138  (2012).

\bibitem{PhysRevLett.106.100403}
M. Hohenadler, T.~C. Lang, and F.~F. Assaad, Phys. Rev. Lett. {\bf 106},
  100403  (2011).

\bibitem{PhysRevLett.107.010401}
S.-L. Yu, X.~C. Xie, and J.-X. Li, Phys. Rev. Lett. {\bf 107},  010401  (2011).

\bibitem{PhysRevB.83.205122}
Y. Yamaji and M. Imada, Phys. Rev. B {\bf 83},  205122  (2011).

\bibitem{PhysRevB.59.2549}
M. Potthoff and W. Nolting, Phys. Rev. B {\bf 59},  2549  (1999).

\bibitem{PhysRevB.79.045130}
H. Ishida and A. Liebsch, Phys. Rev. B {\bf 79},  045130  (2009).

\bibitem{PhysRevLett.72.1545}
M. Caffarel and W. Krauth, Phys. Rev. Lett. {\bf 72},  1545  (1994).

\bibitem{liebsch2011temperature}
A. Liebsch and H. Ishida, Journal of Physics: Condensed Matter {\bf 24},
  053201  (2011).

\bibitem{PhysRevB.81.115134}
G. Borghi, M. Fabrizio, and E. Tosatti, Phys. Rev. B {\bf 81},  115134  (2010).

\bibitem{PhysRevLett.95.066402}
L. de' Medici, A. Georges, G. Kotliar, and S. Biermann, Phys. Rev. Lett. {\bf
  95},  066402  (2005).

\bibitem{PhysRevB.73.245118}
A. Fuhrmann, D. Heilmann, and H. Monien, Phys. Rev. B {\bf 73},  245118
  (2006).

\end{thebibliography}

\end{document}